\documentclass[sigplan]{acmart}
\renewcommand\footnotetextcopyrightpermission[1]{}
\PassOptionsToPackage{hyphens}{url}
\usepackage[shortcuts]{extdash}  %
\usepackage{multicol} %
\usepackage{tabularx, url} %
\usepackage{graphicx} %
\usepackage{enumitem}
\usepackage{hyperref}
\usepackage{xcolor}
\usepackage[many]{tcolorbox}
\usepackage{makecell}
\usepackage{xspace}
\usepackage{amsmath}
\usepackage{xparse}
\usepackage{array}
\usepackage{listings}
\definecolor{Purple}{HTML}{A020F0}
\usepackage{fancyvrb}
\usepackage{verbatimbox,caption,float,lipsum}

\usepackage[capitalize, nameinlink]{cleveref}
\crefformat{section}{#2\S{}#1#3}
\Crefname{section}{Section}{Sections}
\makeatletter
\Crefformat{section}{\Cref@section@name~#2#1#3}
\makeatother

\usepackage{tikz}
\newcommand{\cnumber}[1]{\tikz[baseline=(myanchor.base)] \node[circle,fill=.,inner sep=1pt] (myanchor) {\color{-.}\bfseries\footnotesize #1};}

\newcommand{\bnumber}[1]{\tikz[baseline=(myanchor.base)] \node[circle,draw=black,inner sep=1pt] (myanchor) {\color{black}\bfseries\footnotesize #1};}

\tcbset{
    sharp corners,
    colback = lightgray,
    before skip = 0.2cm,    %
    after skip = 0.5cm      %
}                           %

\newtcolorbox{boxAPI}{
    boxrule = .5pt,
    colframe = black %
}

\newcommand{\sysname}{SYS}
\newcommand{\uswappername}{Swapper}
\newcommand{\sysapi}{\lowercase\expandafter{\sysname}}
\newcommand{\sys}{\textsc{\sysname}\xspace}

\newcommand{\uswapper}{\uswappername\xspace}

\newcommand{\todo}[1]{  }
\newcommand{\note}[1]{  }
\newcommand{\notelv}[1]{ }

\newcommand{\tinyskip}{\vspace{3pt}}

\newcommand{\mypar}[1]{\tinyskip\noindent\textbf{#1.}\xspace}

\title{Flexible Swapping for the Cloud}
\author{Milan Pandurov}\email{milan.pandurov@huawei.com}
\affiliation{%
    \institution{Huawei}
    \city{}
    \country{}
}

\author{Lukas Humbel}\email{lukas.humbel@huawei.com}
\affiliation{%
    \institution{Huawei}
    \city{}
    \country{}
}

\author{Dmitry Sepp}\email{dmitry.sepp@huawei.com}
\affiliation{%
    \institution{Huawei}
    \city{}
    \country{}
}

\author{Adamos Ttofari}\email{adamos.ttofari@huawei.com}
\affiliation{%
    \institution{Huawei}
    \city{}
    \country{}
}

\author{Leon Thomm}\email{leon.thomm@huawei.com}
\affiliation{%
    \institution{Huawei}
    \city{}
    \country{}
}

\author{Do Le Quoc}\email{quoc.do.le@huawei.com}
\affiliation{%
    \institution{Huawei}
    \city{}
    \country{}
}
\author{Siddharth Chandrasekaran}\email{siddharth.chandrasekaran@huawei.com}
\affiliation{%
    \institution{Huawei}
    \city{}
    \country{}
}
\author{Sharan Santhanam}\email{sharan.santhanam@huawei.com}
\affiliation{%
    \institution{Huawei}
    \city{}
    \country{}
}
\author{Chuan Ye}\email{yechuan@huawei.com}
\affiliation{%
    \institution{Huawei}
    \city{}
    \country{}
}
\author{Shai Bergman}\email{shai.aviram.bergman@huawei.com}
\affiliation{%
    \institution{Huawei}
    \city{}
    \country{}
}
\author{Wei Wang}\email{weiwang2@huawei.com}
\affiliation{%
    \institution{Huawei}
    \city{}
    \country{}
}

\author{Sven Lundgren}\email{sven.lundgren@etascale.com}
\affiliation{%
    \institution{Eta Scale}
    \city{}
    \country{}
}

\author{Konstantinos Sagonas}\email{kostis@it.uu.se}
\affiliation{%
    \institution{Uppsala University \& Eta Scale}
    \city{}
    \country{}
}

\author{Alberto Ros}\email{aros@ditec.um.es}
\affiliation{%
    \institution{University of Murcia}
    \city{}
    \country{}
}
\date{May 2024}

\begin{document}
\fancyhf{}
\fancyhf[HR]{} 
\fancyhf[HL]{Flexible Swapping for the Cloud} 
\fancyhf[FC]{\thepage} 

\setlist[itemize]{leftmargin=*}
\renewenvironment{itemize}{
  \begin{list}{\labelitemi}{
      \setlength{\topsep}{0.5ex}
      \setlength{\parsep}{0pt}
      \setlength{\partopsep}{0pt}
      \setlength{\itemsep}{2pt}
      \setlength{\itemindent}{10pt}
      \setlength{\leftmargin}{0pt}}
 }{\end{list}}

\begin{abstract}

Memory has become the primary cost driver in cloud data centers. Yet, a significant portion of memory allocated to VMs in public clouds remains unused. To optimize this resource, "cold" memory can be reclaimed from VMs and stored on slower storage or compressed, enabling memory overcommit. Current overcommit systems rely on general-purpose OS swap mechanisms, which are not optimized for virtualized workloads, leading to missed memory-saving opportunities and ineffective use of optimizations like prefetchers.

This paper introduces a userspace memory management framework designed for VMs. It enables custom policies that have full control over the virtual machines' memory using a simple userspace API, supports huge page-based swapping to satisfy VM performance requirements,
is easy to deploy by leveraging Linux/KVM, and supports zero-copy I/O virtualization with shared VM memory.

 Our evaluation demonstrates that an overcommit system based on our framework outperforms the state-of-the-art solutions on both micro-benchmarks and commonly used cloud workloads. Specifically our implementation outperforms the Linux Kernel baseline implementation by up to $25\%$ while saving a similar amount of memory. We also demonstrate the benefits of custom policies by implementing workload-specific reclaimers and prefetchers that save $10\%$ additional memory, improve performance in a limited memory scenario by $30\%$ over the Linux baseline, and recover faster from hard limit releases.
 \end{abstract}
\maketitle

\section{Introduction}

Memory-intensive applications~\cite{redis, spark, flink, tensorflow} are an integral component of  modern online services. In addition, memory has become the most expensive and sought-after resource in the cloud~\cite{mutlu2013memory}. Thus, to save cost, it is of paramount importance to effectively utilize this expensive resource.

Users typically provision virtual machines' (VM) memory to accommodate for peak loads, deploying large cloud instance types to avoid running out of resources. 
As a result, VMs often under-utilize their memory resources, contributing to the imbalance in memory resource allocation in cloud compute infrastructures. For example, at Google data centers, approximately $30\%$ of server memory is not accessed for minutes and is considered cold~\cite{ruan_aifm_2020}. This suggests that cold memory can be reclaimed and used elsewhere.

To attain flexibility in memory consumption, operating systems (OSes) commonly employ swapping~\cite{lwn_reconsidering_swapping,freebsd_swaps,bergman_znswap_2022}, which conserves memory by paging cold memory to cheaper, slower memory (e.g., disks or remote memory). However, we found such off-the-shelf solutions unsuitable for the cloud, as they do not satisfy one or more of the following requirements:

\begin{itemize}
\item \textit{Flexibility.} A swapping system is heavily dependent on the access characteristics of workloads.
Swapping systems must not affect long running critical workloads and must be \emph{adaptable} (e.g., able to deploy different reclamation algorithms in production).
Furthermore, we do not consider this adaptation to ever be done, as the workloads and access characteristics are constantly changing. Key to achieving this flexibility is the separation of policy and mechanism, to allow modifications to be made to each one independently.

\item \textit{Feedback loop with the control plane.} The cloud control plane requires accurate knowledge of the quantity of cold memory pages to provision more VMs and maximize utilization. Further, it is essential for the control plane to properly configure the swapping system to prioritize VMs and ensure adherence to service level agreements.

\item \textit{Transparency.} Public clouds employ opaque VMs, thus, reclamation mechanisms cannot rely on guest cooperation. 

\end{itemize}

\tinyskip

Current swapping systems are \textit{general purpose}: they identify cold memory pages by using metrics such as relative page access time, access frequency, and type of memory pages (anonymous vs. file-backed). However, they are not tailored to work with VMs: The VMs' backing memories appear to the hypervisor as one large, homogeneous area while in reality many applications are using it, which creates a semantic gap.

In this paper, we introduce a flexible userspace swapping framework for the cloud. It is optimized for virtualization scenarios and offers mechanisms to bridge the semantic gap by inspecting the guest state without guest cooperation.

The developed framework enables efficient exploration of swap system designs by performing all policy decisions in userspace by offering a simple policy API. Further, it disentangles the policy and mechanism of swap systems, implementing both in userspace. This approach facilitates flexibility, such as enabling strict hugepage swapping support, which is imperative for virtualization, and is unavailable in Linux.

Our framework adds the ability to integrate cloud control-plane functionality into the swap system, such as informing the control plane about the number of cold memory pages. This ability in isolation can be achieved with existing policies and mechanisms~\cite{nutanix}. However, this method has several shortcomings, such as redundant cold page estimation, which leads to sub-optimal decisions, as we show in~\cref{sec:eval-kernel-baseline}.

Our system supports efficient memory reclamation by offering virtualization specific optimizations for opaque VMs. Unlike general-purpose swap systems, it addresses the need for transparency by leveraging VM contextual information for guest VMs, while avoiding reliance on guest cooperation.

Our work makes the following contributions: 
\begin{itemize}
    \item An examination of the trade-offs between performance, data and information granularity in VMs when utilizing 2MB versus 4kB pages with swapping mechanisms (\cref{sec:anal-hugepages}).
    
    \item An in-depth analysis of swap system overheads in the virtualization and cloud vendor scenarios (\cref{ssec:access-patterns-degrade} and \cref{ssec:anal-ept-scan-expensive}).

    \item %
    Design (\cref{sec:design}) and implementation (\cref{sec:implementation}) in userspace of a memory overcommit swap system targeted at opaque VMs.

    \item An analysis of the opportunities presented when performing swap in userspace, overheads, and their mitigation (\cref{sec:eval}). 

    \item An extensive evaluation of the developed system using synthetic and real-world workloads. It outperforms Linux-based swapping by $25\%$ under best-effort cold memory reclamation, and by $30\%$ in limited memory scenarios. Flexibility is demonstrated in a workload-specific reclamation that saves an additional $10\%$ of memory with only small performance degradation.
\end{itemize}
\section{Background and Motivation}
\label{sec:motivation-hugepages}

\mypar{Nested paging}
Most hypervisors use a form of hardware assisted nested paging.
In this model, the guest OS is in charge of the guest-virtual to guest-physical address translation (\textit{GVA} $\rightarrow$ \textit{GPA}), given by the \textit{CR3} register.
The hypervisor maintains a second translation in the extended page-table (EPT)~\cite{intel_sdm}, which translates from GPA to host-physical addresses (\textit{GPA} $\rightarrow$ \textit{HPA}). Like a regular page table, the EPT contains access- and dirty-bits that can be leveraged to infer which memory is accessed.

In the case of type-2 hypervisors, such as the Kernel Virtual Machine (KVM)~\cite{kvm} on Linux, a control and device emulation process runs in user-space (like QEMU~\cite{qemu}). Given its nature as userspace process, it can not access \textit{GPA}s directly, rather it uses a dedicated mapping to access guest memory using host-virtual addresses (HVA).

\mypar{Linux Hugepages} Hugepages are commonly used to reduce TLB pressure, shorten page walks~\cite{agbarya2018memomania, fuerst2022memory, panwar2019hawkeye} and improve performance~\cite{manocha2023architectural}. A hugepage TLB entry covers $512\times$ more memory than a regular entry. As a result, using hugepages can significantly improve TLB hit rates.

Linux applications can use 2MB pages using Transparent Huge Pages (THP)~\cite{thp} or HugeTLB Pages~\cite{hugetlb}. While THP requires no application modification, HugeTLB requires explicit allocation.
THP can be split into 4kB pages by Linux (e.g., to avoid fragmentation).
On the other hand, HugeTLB are never split into 4kB pages and users can rely on the memory being backed by hugepages all the time, %
therefore, making it a popular choice for cloud virtualization hosts.

\mypar{Linux Swapping and Control Groups}
The kernel identifies old pages using two LRU lists, pages are promoted upon page fault, and  demoted when under memory pressure~\cite{gorman_chapter_nodate}.
HugeTLB swapping is not supported by Linux, and THP backed memory will be split into 4kB pages upon page fault.
By default, the Linux kernel only \textit{reactively} swaps out under memory pressure. Control Groups~\cite{tejun_linux_2015} (\textit{cgroups}) can be memory limited to force a certain amount of memory reclamation and swapping.

KVM supports \textit{asynchronous page faults}, a mechanism that notifies the VM if a page fault occurred before the page fault is handled.
The rationale is if the guest also supports this mechanism, it can schedule another context which is (presumably) paged in. Using this mechanism it is possible to have multiple page faults in-flight even with just one \textit{vCPU}.

\section{Analysis}

\subsection{The case for hugepage swapping}
\label{sec:anal-hugepages}

\begin{figure}
    \centering
    \includegraphics[trim={0em 0.8em 0em 0em},clip,width=0.9\columnwidth]{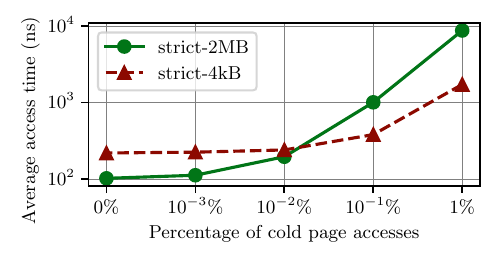}
    \caption{Average access latency (ns) with varying percentages of cold-page accesses}
    \label{fig:mb-case-for-hugepages}
\end{figure}

While hugepages are preferable for virtualization, the benefits are less clear if we consider a swapping system, as there is a trade-off between the advantages of hugepages in-memory and the faster 4kB swap operations. In this subsection we investigate, using a microbenchmark, the trade-off between \textit{strict-4kB} systems, where both memory and backing store utilize 4kB pages, and \textit{strict-2MB} systems, which utilize 2MB pages. Mixed systems that split pages on swap-in/swap-out are also conceivable, but we discuss them later and instead focus on the extremes of the design space.

We use a microbenchmark in a VM that performs uniformly random memory accesses to two regions in the memory address space: the first consists of resident memory pages and the second is to swapped-out memory pages which reside in an SSD backing store, triggering a swap-in operation. We gradually adjust the ratio of random memory accesses to the resident memory pages and swapped-out pages (\textit{cold-page access ratio}). We then measure the average memory access latencies for both the \emph{strict-2MB} and \emph{strict-4kB}. In both cases, the size of the swapped-in memory region is large enough to ensure near-100\% TLB miss ratio.

\Cref{fig:mb-case-for-hugepages} shows the results. It is apparent that there are two different effects at the extremes of the cold-page access ratio values. At low ratio values, where swap-in operations seldomly occur, we observe the benefits of hugepage access over 4kB due to the shortened page-table traversal overheads. At higher ratio values, 4kB exhibits shorter average memory access latency due to the smaller granularity pages leading to less swapped-in data on each page fault. We notice that once we pass the \textit{2MB/4kB break-even} of $0.01\%$, \textit{strict-2MB} usage starts to be faster than the \textit{strict-4kB} ones.

While this ratio is important, because it predicts the application performance of a partially swapped out application, it is not the only factor to consider when deciding on the granularity. For instance, opting for 2MB pages might result in less cold memory being identified due to hotness fragmentation~\cite{bergman2022reconsidering}. Conversely, using 2MB pages accelerates page-table scans owing to smaller page table sizes. 2MB pages can reduce the cold-access percentage on workloads that have sufficient locality and reuse many 4kB parts of a 2MB page. We investigate these factors later (\cref{ssec:anal-ept-scan-expensive} and~\cref{ssec:eval-softlimit}).

\subsection{Access patterns degrade under virtualization}
\label{ssec:access-patterns-degrade}
A hypervisor dealing with opaque VMs only sees the GPA space directly. GPAs can be trivially converted to HVAs, but it is not possible to obtain GVA ranges. GVA $\rightarrow$ GPA translation is under the guest's control, can be arbitrary, and changes with each context switch. Spatial patterns (e.g., neighboring pages accessed frequently together) are not necessarily visible in the (easy observable) GPA/HVA address space.

To illustrate this effect, we launch 4GB VM and warm it up by running a random memory access process for 1 second to age the memory subsystem on the VM. We then run a microbenchmark that allocates 3GB of memory, accesses only the first half of the memory for one minute, and then switches to the second half of the memory. We sample the access bit in a short interval and plot the pages accessed. 

\begin{figure}
    \centering
    \includegraphics[trim={0em 1.15em 0em 0.8em},clip,width=0.9\columnwidth]{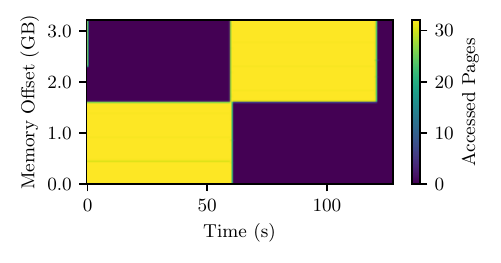}
    \includegraphics[trim={0em 1.15em 0em 0.8em},clip,width=0.9\columnwidth]{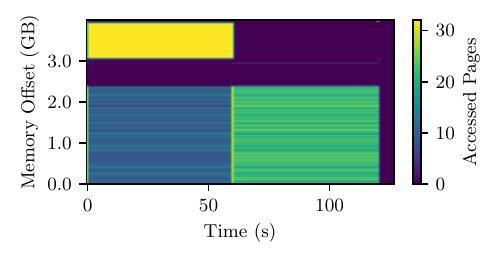}
    \caption{Page access pattern of a $50\%/50\%$ alternating workload  measured directly (up) and under virtualization (down)}
    \label{fig:virt-degrades-access}
\end{figure}

\Cref{fig:virt-degrades-access} shows the resulting accesses heatmap as measured from inside the VM and from the hypervisor. The scrambling under virtualization is due to the way the guest operating system allocates pages and sets-up page tables which is outside the control of the hypervisor. This effect does not only appear in our setup (QEMU with KVM), but any hypervisor using nested paging. 

This observation leads to a key challenge: when making spatial predictions about the memory access, we must not rely on the guest physical address space, rather must we inspect the current guest application address space (\cref{impl:gva-to-hva}). Furthermore, this observation has consequences for the applicability of existing techniques (\cref{ssec:eval-linearpf}).

\subsection{EPT scanning affects guest VM performance}
\label{ssec:anal-ept-scan-expensive}

The process of scanning and zeroing the access bits to determine cold memory pages incurs system overheads~\cite{park_daos}.

\begin{figure}
    \centering
    \includegraphics[trim={0em 1em 0em 0.8em},clip,width=0.9\columnwidth]{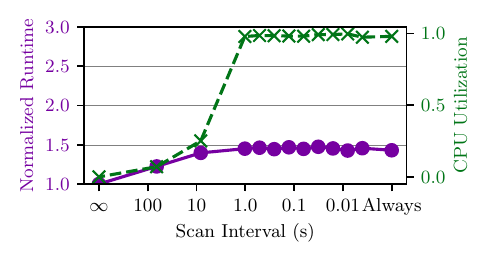}
    \includegraphics[trim={0em 1em 0em 0.8em},clip,width=0.9\columnwidth]{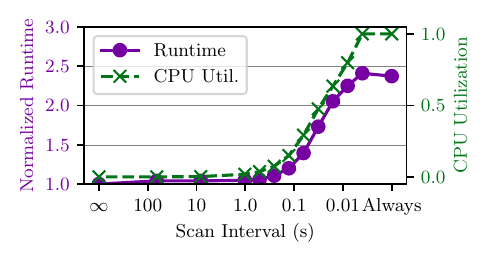}
    \caption{Direct (\textit{\% CPU}) and indirect (\textit{runtime}) costs
    of increasing the EPT scan frequency with 4kB (up) and 2MB pages (down).}
    \label{fig:ept-scan-cost}
\end{figure}

We perform a scan of the access bits and zero them without flushing the TLB while running a sequential read-only memory scan on a different CPU core.
EPT scanning comes with \textit{direct} cost caused by the CPU utilization of the scanning process and an \textit{indirect} cost by slowing down the application, caused by partial-walk-caches flushed~\cite{intel_sdm} on clearing the access bits.
Thus, \cref{fig:ept-scan-cost} shows the effect of decreasing the EPT scan interval on the workloads runtime and on the CPU utilization of the scanning core.
Both the direct and the indirect cost increase as the scan interval is decreased. However, a high scan interval means losing reactivity, as we have to wait for the next scan interval, and precision, because more accesses will be grouped together.

\subsection{DAMON}
\label{ssec:damon}
The data access monitor subsystem DAMON~\cite{park_daos} is a Linux proactive reclaimer, that also tackles the cost of access bit scanning (see \cref{ssec:anal-ept-scan-expensive}). To lower this cost, DAMON relies on dynamically constructed ``regions'', consisting of consecutive pages, which are randomly sampled.
If a whole region reaches a configurable age, DAMON can perform a set of \texttt{madvise} actions on the region, such as page-out.

The current implementation of DAMON does not support scanning the EPT, but only regular processes' virtual and physical address spaces. While it is technically possible to implement an EPT scanning backend for DAMON, we do not expect the region based sampling approach to work well given our observation of access pattern degradation (see \cref{ssec:access-patterns-degrade}). Additionally, other works \cite{10.1145/3627703.3650075, memtis} report a lack of precision even without virtualization. %

\section{Design}\label{sec:design}

Our userspace framework is designed for building memory overcommit systems supporting flexible per-VM swapping policies. We aim to achieve the following design goals:

\begin{itemize}
  \item \emph{Flexibility.} Effective memory overcommit requires sophisticated algorithms, which are difficult to develop in kernel space, due to complexity, and lack of memory- and performance-isolation. The developed system addresses this by implementing such algorithms in user space, running them isolated off the critical path and providing a safe \textit{policy-API}. Together, these allow rapid policy development. The policy-API abstracts hardware features such as page size, architecture, and storage medium and provides a unified mechanism to obtain memory access bitmaps. The API is safe in the sense that an API client is not able to corrupt guest memory or violate memory limits.

  \item \emph{Built for the virtualization in the cloud.} As described in \cref{ssec:access-patterns-degrade}, developing effective, or even simply \emph{correct}, swap related algorithms require visibility into the guest VM. The developed system provides low overhead VM introspection mechanisms. Further, we aim for its deployment in a cloud production environment. Thus, it must integrate with existing stacks, like accelerated networking using Open vSwitch (OVS~\cite{ovs}) and storage stacks (SPDK vhost~\cite{spdk}) relying on direct memory access (\textit{DMA}). To accelerate adoption, our system must not require any modifications to the guest OS.

\end{itemize}

\subsection{Design overview}
\label{design:overview}
\begin{figure}
    \centering
    \includegraphics[width=.48\textwidth]{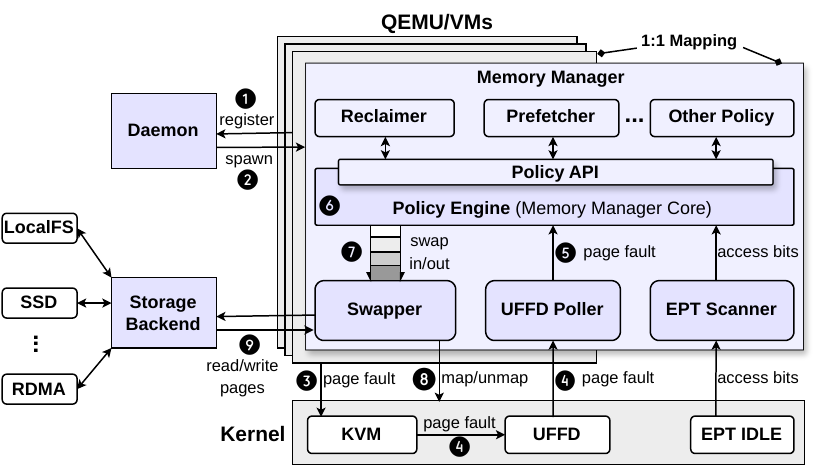}
    \caption{system overview}
    \label{fig:uswap-overview}
\end{figure}

Figure~\ref{fig:uswap-overview} shows the high-level architecture of the system. It consists of several processes, including the daemon, the memory manager (MM), and the Storage Backend. Daemon is launched at system startup and is responsible for spawning, configuring, and managing MM processes for each virtual machine (VM). During the boot phase, each VM process (QEMU) informs daemon \cnumber{1} with the desired page size and service level agreement. Based on these, daemon determines a memory management configuration and launches it to manage memory \cnumber{2}. The Storage Backend is a standalone application that provides input/output services to multiple MMs.
MM runs the Policy Engine, mandatory modules (swapper, UFFD poller, EPT scanner), and optional modules.
Modules that run swapping algorithms like reclaimers and prefetchers are optional and we refer to them as \textit{policies}.

Memory manager modules can export a set of parameters that can be controlled during runtime through the MM API. Modules can register a callback function that gets invoked when an external application reads or writes to the parameters.

We describe the high-level workflow of MM using two different tasks: handling a page fault and reclaiming unused memory.

\mypar{Life of a page fault}
When a VM accesses swapped out data, it causes an EPT violation in the hypervisor \cnumber{3}. Once the EPT violation has been raised, the faulting address is 
passed to the Linux MM subsystem, which in turn notices this area is backed by userfaultfd, Linux's mechanism for handling page faults in userspace (UFFD~\cite{uffd}). The event is then passed to the user space \cnumber{4}.
The UFFD Poller listens to such events and propagates them to the Policy Engine \cnumber{5}. The Policy Engine first verifies that servicing the newly received page does not result in a memory limit overrun \cnumber{6}. Assuming that this memory limit is not exceeded, the Policy Engine enqueues the necessary work in the \uswapper queue and notifies policies asynchronously. The \uswapper module runs multiple worker threads, and one of them picks up the new work from the queue \cnumber{7}. 
The \uswapper worker thread first contacts the storage backend via a shared memory communication channel, and requests a page to be loaded into memory \cnumber{9}. The storage backend reads data from the storage device and populates the VM's memory with correct page content and signals completion to the \uswapper thread. The \uswapper thread then instructs the kernel to map \cnumber{8} the faulting page into the VM's address space and complete the UFFD event. This resolves the outstanding EPT violation and the VM continues its normal execution. 

\mypar{Reclaiming unused memory}
To identify cold pages, MM's modules receive access bitmaps from the EPT scanner periodically. The EPT scanner uses a custom kernel module (see \cref{ssec:impl-ept-scan}) to read and clear the access bit and generate a bitmap. The reclaimer (an optional module) subscribes to these events and utilizes them
to determine which pages are cold. Upon identification of cold pages, the reclaimer requests pages to be reclaimed through the policy API.
Once the Policy Engine receives this request, it will verify that the page is indeed swapped in and send a swap-out request to \uswapper. For swap-out operations, the \uswapper thread will instruct the kernel to unmap the page from all of the clients.
Afterwards, \uswapper requests the storage backend to store the page's content to the underlying disk and finally deallocates the memory from the kernel.

\subsection{Memory Manager (MM)}
\label{design:memory_manager}

Memory manager is an userspace process taking care of paging and memory reclamation for an associated VM. It consists of the Policy Engine and integrates modules responsible for scanning memory and handling userspace faults.

For enhanced userspace page fault handling, memory manager provides the policies in the Policy Engine of additional contextual VM information. To avoid redundant swap operations and enforce task priorities during swap operations, the Policy Engine schedules work to \uswapper through special priority queue -- \uswapper queue. Each MM relies on a Storage Backend application to save and restore pages from cold storage.

\mypar{Lightweight VM introspection} MM attempts to bridge the semantic gap between the policies operating in the hypervisor context and applications running inside the guest VM while being fully transparent to the VM execution flow. 
MM provides additional context associated with the page fault to the policies such as VM-specific registers which can be used to build more accurate policies. For example, a policy can differentiate between guest applications by examining the VM's control registers. It also provides a light-weight translation component between GVA and HVA. This is useful as regular application memory access patterns are generally only observable in GVA space, while the hypervisor and rest of the MM internally only work with HVA.

\mypar{\uswapper queues} \uswapper queues are a priority-queue pair that can hold different requests types: page fault, reclaim, and prefetch requests. The \uswapper queue is designed to avoid redundant I/O operations and enables prioritisation of different request types, such as prioritizing page fault over prefetch requests, without violating memory limits.

As the Policy Engine can enqueue requests faster than \uswapper can process them, multiple conflicting requests may accumulate in the queue. For instance, a policy might decide to swap out a page while a request to swap in the same page is already queued, resulting in redundant operations that waste time and I/O resources. To address this issue, the \uswapper queue is designed to enqueue only an indication of the pages that require action, rather than specifying an explicit operation (i.e., swap-in). \uswapper will then dequeue the page, determine the necessary state of the page, and perform the required actions to achieve the desired state. If the page is in its correct state, no action will be taken.

\subsection{Policies}
\label{des:vm-introspection}
\label{ssec:design-policy}
\begin{table}
\centering
\footnotesize
\begin{tabular}{ | p{11em} | p{16em} | }
 \hline
 Function Name  & Description \\ 
 \hline
 \texttt{reclaim(addr)} & Request page address to reclaim \\  
 \texttt{prefetch(addr)} & Request page address to prefetch \\
 \texttt{on\_event(evt\_type, cb)}  & Register event callback: page fault, memory limit change, swap out, swap in. \\
 \texttt{gva\_to\_hva(gva, cr3)}  & Convert from GVA to HVA \\
 \texttt{scan\_ept(scan\_int, cb)}  & Scan VM's EPT at regular intervals and invoke the callback with access bitmap \\
 \texttt{get\_page\_state(addr)}  & Get current state of the page (swapped IN or OUT) \\
 \texttt{get\_memory\_limit()}  & Get current memory limit \\
 \texttt{get\_memory\_usage()}  & Get current memory usage \\
 \texttt{get\_pf\_count()}  & Get current page fault count \\
 \texttt{register\_parameter(name, read\_cb, write\_cb)}  & Registers a read \& write callbacks for a parameter. Parameter is accessible through MM-API. \\
 \hline
\end{tabular}
\caption{Policy API}
\label{tab:api}
\end{table}

Policies are how our system achieves flexibility: they are optional software components that can subscribe to events (EPT scans, page faults, etc.) and instruct the Policy Engine to prefetch or reclaim pages via the policy API, described in~\cref{tab:api}. MM can run an arbitrary number of policies depending on the VM's configuration and desired SLO.

A policy using the policy API cannot compromise the correctness (e.g., corrupting guest memory) nor impact the page fault performance, except forced reclaim under memory limit. We achieve this by executing policies in a separate thread and providing them with a separate event queue.

\mypar{Policy Engine} The Policy Engine synchronizes requests coming from policies and page faults coming from the UFFD poller and inserts necessary work to \uswapper queue for processing. As outlined in \cref{design:overview}, \uswapper is consuming requests from this queue and processing them asynchronously. Each time the Policy Engine adds a request to the \uswapper queue it will adjust memory usage, for example on swap-out it will decrease memory usage and increase it for swap-in. In case an incoming request would violate the memory limit, the Policy Engine will either drop the request (in case of a prefetch request) or invoke a special memory limit reclaimer to determine which page to swap out. The Policy Engine will make sure that \uswapper queue has the correct ratio of swap-in and swap-out requests i.e. when all requests from the queue get processed, the memory limit won't be exceeded.

\mypar{Forced memory reclamation} When a page fault occurs while the memory usage is at the limit, the Policy Engine must force a reclamation. The memory limit reclaimer is a special policy that is invoked synchronously to determine the target page for reclamation. This reclaimer needs to make this decision quickly since it lies on the page fault processing path. Our default implementation is a LRU-based reclaimer.

\mypar{Example} We demonstrate how using the policy APIs can simplify the implementation of smart policies by designing an application-aware next page prefetcher. In contrast to the simple next (physical) page prefetcher, the application-aware one understands the guest application's address space and can fetch a physical page that corresponds to the next page in the virtual address space. Most of the heavy lifting is provided by the VM introspection APIs, described in \cref{design:memory_manager}. The following example assumes x86 architecture where CR3 holds the page-table pointer.

\begin{small}
\begin{verbatim}
void on_page_fault(page, cr3, gva) {
    if (!cr3 || !gva) {
        // Page fault has no associated 
        // CR3 or GVA info. Don't prefetch.
        return;
    }
    next_gva = gva + page.size();
    next_hva = SYS.gva_to_hva(next_gva, cr3);
    if (!next_hva) {
        // GVA to HVA can fail, don't prefetch.
        return;
    }
    SYS.prefetch(next_hva);
}
\end{verbatim}
\end{small}

The example code shows the page fault event callback function. Our system will attach additional context registers like the GVA and CR3. The example policy will then increment the GVA by the page size, convert it to HVA, and issue a prefetch. Converting from GVA to HVA is necessary because the hypervisor only understands HVAs. Despite the prefetcher's straightforward implementation, it shows promising results as described in \cref{ssec:eval-linearpf}.

\subsection{Storage Backend}
\label{sec:spdk-storage-backend}
The Storage Backend is a single userspace process providing all swap I/O services to MMs. The component receives requests from the swapper workers and subsequently exchanges data with one or more connected I/O devices. It is able to multiplex requests coming from different MMs, and save and restore pages from cold storage mediums such as NVMe disk, RDMA, etc. 

The Storage Backend is designed to easily interact with different I/O-optimized userspace frameworks, such as SPDK, DPDK, etc.

\section{Implementation}\label{sec:implementation}

\subsection{Linux integration}
\label{sec:linux-integration}

\begin{figure}
    \centering
    \includegraphics[width=.48\textwidth]{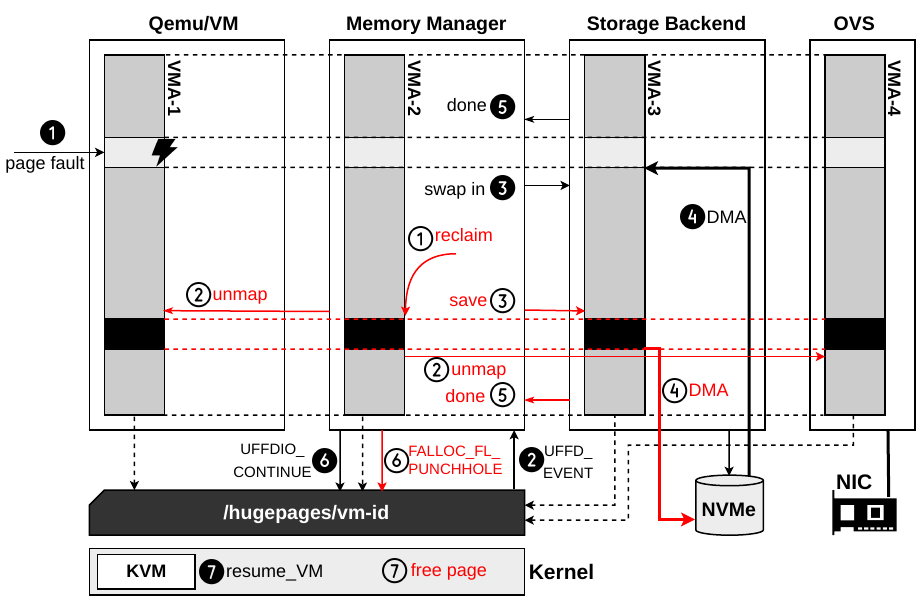}
    \caption{Swap-in (in black) and swap-out (in red) using shared memory}
    \label{fig:uswap-shared-memory}
\end{figure}

Our system is built on top of Linux shared memory, UFFD and a handful of additional memory management kernel syscalls to implement swap memory management in userspace. \cref{fig:uswap-shared-memory} shows  processes and their memory mappings in Linux.  For easier integration in client applications like QEMU and OVS, our system provides a shared library that exposes mmap-like functions for mapping memory. Allocating new memory through the shared library will internally create a memory-backed file, map it to the application's address space, register newly mapped memory to UFFD and send UFFD information to MM. MM and storage backend will both map the shared memory file in their own address space. Additional clients like OVS can also use the shared library to mmap the managed memory.

\mypar{\emph{Swap in path}} When clients like QEMU or OVS access memory that is not mapped in their address space, a page fault \cnumber{1} will be triggered and a UFFD page fault event is sent to MM \cnumber{2}. When the storage backend receives a request to swap in a page from MM, it instructs the NVMe disk to DMA the page's content directly inside the VM's memory \cnumber{4}. This is possible as the storage backend has the VM's memory mapped in its own address space as well. Once the memory content is populated, MM will issue \texttt{UFFDIO\_CONTINUE} ioctl \cnumber{6}, which will effectively resolve the page fault and resume the client's execution.

\mypar{\emph{Zero-page pool}}
To avoid having to zero a 2MB page, which lasts around $100us$, on the critical swap-in path, we introduce a zero-page pool for 2MB pages and use idle time to refill the pool.

\mypar{\emph{Swap out path}} When MM decides to swap out a page \bnumber{1}, it will first unmap the page from each of the clients \bnumber{2} by issuing  \texttt{MADVISE\_DONTNEED} through process\_madvise syscall. Once this step is finished, clients accessing memory will trigger page fault which ensures that memory cannot be modified wihout MM's knowledge. It will request storage backend to save the page \bnumber{3} and it will DMA page's content to NVMe disk \bnumber{4} and finally MM will issue \texttt{FALLOC\_FL\_PUNCHOLE} syscall \bnumber{6} on a backing memory file, effectively freeing used memory.

\subsection{VM introspection}
\label{impl:vmcs-info-propagation}
\mypar{Enhancing page fault information} 
As described in \cref{des:vm-introspection}, our system adds additional guest information to the page fault from the VM's registers at the time of fault. To avoid passing this information through Linux's memory subsystem, we add a ring buffer between kernel and MM. At the time of the EPT violation, before passing the page fault to the kernel's MM subsystem, KVM is modified to copy a set of registers from the virtual machine control structure (VMCS) into this ring-buffer. Currently page directory base pointer (PDBP), instruction pointer (IP), and guest linear address (GVA). 
Once the UFFD fault event is received by MM, it will consume information from the ring buffer, find associated context registers, and attach it to the page fault event, which will be received later by policies.

\mypar{Translating GVA to HVA}
\label{impl:gva-to-hva}
Smart policies can use this additional page fault information to predict which pages to prefetch or reclaim in the GVA space. However, to issue such a request, the policies need to convert the predicted GVA to HVA such that the system can understand it. The conversion between GVA and HVA is done by walking the guest page tables for a given PDBP and GVA. Our framework exposes an API for doing this conversion. Internally, it forwards the request from MM to the QEMU process.
QEMU will perform the page table walk in a separate thread and return the desired address to the policy. If the guest page tables are outdated or do not yet have a mapping for a given virtual address, the translation process may not succeed and return an error.
In our experiments, we have noticed only a small fraction of all translations do not succeed, and can be ignored.

\subsection{SPDK storage backend implementation}

In the current implementation we target NVMe as swap medium since it is easily available in cloud hosts. When a new MM is created it will establish a communication channel with the storage backend and the storage backend will map the VM's memory into its address space, as outlined in \cref{sec:linux-integration}. MM and the storage backend are communicating using a lock free queue in which swapper threads insert I/O requests and go to sleep waiting on a semaphore. On the other side, the storage backend polls for new requests and processes them. The SPDK storage backend, the default storage backend, is built using the SPDK libraries used for high-performance storage access and is using one physical CPU core to poll the SSD. Once the SPDK storage receives a new request it will program NVMe's DMA engine to read or write page data, for 2MB pages this is achieved in zero-copy fashion by programming DMA to access the VMs memory directly.
In the 4kB case, the data is copied into a bounce buffer, since SPDK does not support 4kB for DMA. Once the DMA transaction is finished, the SPDK storage backend will identify which page it was associated with and wake up the correct swapper thread in MM.

\mypar{Addressing the SPDK overhead} The CPU core that is occupied by the SPDK application cannot be used for anything else. Sacrificing one CPU core for the sole purpose of handling SWAP I/O is not cost effective. Fortunately, many cloud host setups are already using SPKD/DPDK applications for different purposes e.g. implementing accelerated networking and disk emulation. System provides a storage backend library that can be easily integrated inside an existing DPDK/SPDK application and reuse existing CPU polling time. This way, system adds minimal overhead to the overall CPU load.

\subsection{Memory reclamation}
\label{ssec:impl-ept-scan}

\mypar{EPT scanner} Scanning the EPT is only possible from kernel space. Hence, we add a Linux kernel module based on the Intel memory-optimizer~\cite{intel_memory_optimizer}, which exposes access-bitmaps of KVM-based VMs to userspace. The EPT scanner of MM is the userspace counterpart that aggregates policy requests for scans, and reads and forwards this bitmap periodically to policies.

Given our observation in \cref{ssec:anal-ept-scan-expensive} and \cref{ssec:access-patterns-degrade}, we do not support hierarchical access bit tracking or sampling non-leaf page-table entries. Rather we allow policies to dynamically adjust the scanning interval.

To support Virtual I/O devices (VIRTIO~\cite{virtio_spec}), it is necessary to scan QEMU's page table too, as in some extreme cases, like serving static files, as little as 50\% of the working set is accessed from within the VM. Thus, we extend the module to force an additional scan of QEMU's page table.

\mypar{Default memory reclaimer}
\label{ssec:impl-sys-default-reclaimer}
The default proactive reclaimer and consumer of access bitmaps is the \textit{dt-reclaimer} policy, based on a recent publication~\cite{google_far_memory}. The dt-reclaimer scans the EPT in a configurable
interval (defaults to $60s$) and maintains a history of access bitmaps.
It considers pages not appearing in the last \textit{threshold} bitmaps as cold and reclaims them.

To determine the \textit{threshold}, it maintains per-page access-distance histograms which are updated on each interval.
The histograms of all the pages are then used to calculate a \textit{proposed threshold}, such that at most a $X\%$ of the working set is predicted to fault during the next interval. $X$ is the \textit{target promotion rate} and is used to configure the aggressivity.
By default, following the publication we set this value to $2\%$. To avoid threshold fluctuations, the final threshold is smoothed out from the current and past proposed thresholds.

\subsection{Achieving zero-copy I/O}

To achieve high network and disk I/O performance inside VMs, cloud providers often deploy zero-copy solutions for interaction between devices and VMs. For example, networking is usually handled by the OVS which uses DPDK to accelerate I/O. In this case, the OVS application has access to the VM's memory and can instruct devices to DMA data directly out of guest buffers. Devices must be prevented from DMAing to swapped-out pages. Hardware solutions for this problem exist~\cite{pf_nic} but they are not widely available in data centres.

\mypar{Page locking} To overcome this challenge system introduces a lightweight page locking mechanism that can be utilized by applications like OVS to prevent pages from being swapped out while they are being used as DMA source/destination buffers. The core of the locking mechanism is a shared memory region between a client and MM which contains a page-lock bitmap, indicating which pages are locked. Locking a page is a two-step process, the first client will atomically set the page's bit in the locked bitmap and afterwards read from that page to trigger a swap-in, in case the page was swapped out. Once the read succeeds and the lock bit is set, MM will not swap out the page. Once the DMA is finished, clients unlock the page by atomically resetting the bit. Our shared library provides convenient APIs for locking pages.

\section{Evaluation}
\label{sec:eval}

In this section, we methodically evaluate systems components and its core abilities. In \cref{sec:eval-mb-kernel-compare} we compare the framework's mechanism to Linux kernel swap and in \cref{ssec:eval-mb-reclaim} we show how systems's working set size approximation approaches the known working set size of a microbenchmark. 

In \cref{sec:anal-hugepages} we established that  \textit{strict-2MB} swapping can be faster under rare cold-page accesses. In \cref{ssec:eval-softlimit} we illustrate how reclamation stays below this threshold and how the larger granularity of 2MB affects reclamation.

We then perform two application-level comparisons to a Linux-based baseline implementation: In \cref{sec:eval-kernel-baseline} we compare performance and memory saved under proactive reclamation, while in \cref{ssec:eval-reuse} we compare performance under a heavy swapping scenario (thrashing).

Finally we demonstrate the flexibility of the developed system in four example variations of the default policies: An improved memory limit reclaimer~\cref{ssec:eval-reuse}, a simple prefetcher~\cref{ssec:eval-linearpf}, a faster reclaimer~\cref{ssec:eval-aggressive} and faster recovery from a thrashing memory limit~\cref{ssec:eval-ws-restore}. 

\mypar{Machine setup} 
We performed all evaluations on a server with two Intel Xeon Gold 6226 Cascade Lake sockets. Power save and hyper-threading are disabled in the BIOS. 384 GB DRAM are installed, half on each NUMA node.
We utilize a dedicated server grade 4 TB Intel® D7-P5510~\cite{intel_ssd} SSD as a swap target, connected using 4x PCIe Gen3.

\mypar{Benchmark setup}
We run the custom Linux kernel 5.14  described earlier~\cref{sec:implementation} as our hypervisor. When we compare against Linux, we use the default value $3$ for \texttt{vm.page-cluster} and asynchronous page faults are enabled.

Our default VM uses 8 vCPUs and 128GB of RAM. We pin the QEMU vCPUs individually to physical CPUs on the same NUMA node as the SSD. Baseline experiments use the same CPU set.
In baseline and the developed system experiments, we force memory to be used from the SSD's NUMA node by only allocating HugeTLB pages from that NUMA node. In the case of Linux swapping experiments, we rely on Linux's first-touch policy to obtain memory from that NUMA node.

\mypar{Comparing to Linux swapping}
When using Linux swapping, we rely on CGroups to provide memory usage and to force swapping. Unlike the developed system, CGroups do not allow isolating VM memory from QEMU memory. We compensate for this fact by adding a constant whenever setting the limit. The constant is determined as the maximum overhead by measuring a workload run of QEMU with HugeTLB, which allows separating between VM and QEMU memory. Since the constant equals to QEMU's maximum consumption, the Linux version usually has \emph{more} memory available than the developed system, due to QEMU's variance in memory usage.

\mypar{Comparing memory saved}
To calculate the memory saved between two executions, we divide the faster runtime into $5s$ buckets and align these buckets with the slower runs, using benchmark-specific synchronization points. Thus, the slower run's buckets might be longer than $5s$. We then average the relative memory over the buckets.

\subsection{Mechanism performance}
\label{sec:eval-mb-kernel-compare}

We evaluate the page fault handling latency and overall throughput of the developed system and Linux swap mechanisms by executing a microbenchmark in a VM. The benchmark initializes a memory region, instructs the hypervisor to swap out the entire memory, and then initiates a workload that performs random, page-aligned memory accesses.

We disable prefetching, read-ahead, and asynchronous page faults for both developed system and the Linux kernel. These mechanisms increase the latency of single page faults in the hope of amortizing it over future page faults.

\begin{figure}
    \centering
    \includegraphics[trim={0em 0.7em 0em .5em},clip,width=0.9\columnwidth]{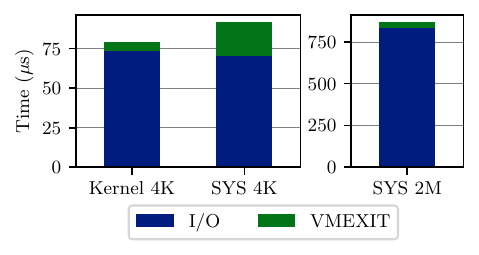}
    \caption{Time spent swapping, the developed system vs kernel}
    \label{fig:mb-vmexit-breakdown}
\end{figure}

\mypar{Page fault latency}
We use host-side tracing to break down the cost of a page fault into software-induced overhead (\textit{VM\-EXIT}) and disk I/O. \cref{fig:mb-vmexit-breakdown} shows this cost breakdown.

Due to additional context switches of userspace swap, our system incur additional overheads. While the \textit{VMEXIT} cost grows from $6us$ to $22us$ comparing 4k to Kernel-4k, the total latency, including I/O, increases by only $12us (13\%)$.
The relative contribution of \textit{VMEXIT} in 2M swapping is lowest ($4.2\%$), even lower than the relative contribution in Kernel-4k ($7.8\%$). Nevertheless, the cost of a 2M page fault is $11\times$ higher than a Kernel-4k page fault, while loading $512\times$ as much data.

\begin{figure}
    \centering
    \includegraphics[trim={0em 0.7em 0em .4em},clip,width=0.9\columnwidth]{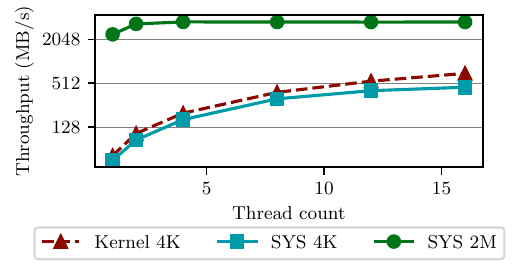}
    \caption{Throughput of swapping related I/O}
    \label{fig:mb-swap-throughput}
\end{figure}

\mypar{Throughput}
We compare the scalability of the kernel and the developed system by increasing the $vCPU$ count and workload threads. 

We show the throughput in \cref{fig:mb-swap-throughput}.
For the page size of 2MB, the developed system saturates the bandwidth of the system with 2 swapper threads. 4kB page size scenarios demonstrate comparable throughput for both developed system and kernel.

We independently verify the maximum throughput achievable in our machine with \textit{fio}~\cite{fio}
and obtain a maximum throughput of around 2.6GB/s, which matches the maximum throughput 2M system configuration achieves. Our machine's throughput is limited by the PCIe v3 bus, explaining why we do not obtain the vendor specified throughput~\cite{intel_ssd}.

\subsection{Working set size estimation}
\label{ssec:eval-mb-reclaim}

We demonstrate that developed system's working set estimation approaches the effective working set. This is done by running a synthetic workload in a VM with a known, varying working set and comparing the value against the reported values from the system.

\begin{figure}
    \centering
    \includegraphics[trim={0em 0em 0em .5em},clip,width=0.9\columnwidth]{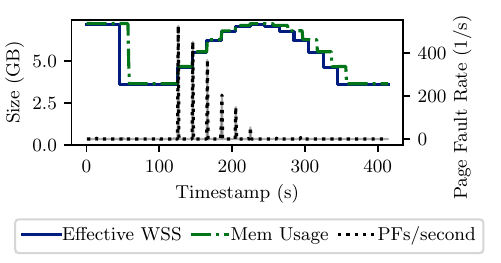}
    \caption{Effective WSS, memory usage, and page fault rate over time}
    \label{fig:mb-wss-pf-mem-usage}
\end{figure}

\Cref{fig:mb-wss-pf-mem-usage} demonstrates effective WSS values reported directly by the microbenchmark and the memory usage and page fault rate reported by MM.
We see that the reclaimer can precisely estimate the actual WSS and the memory consumption adjusts to this estimate in a timely manner.

\subsection{Performance retention and memory saving}
\label{ssec:eval-softlimit}

We evaluate the memory savings and performance impact of 2M and 4k compared to a non-swapping baseline. The Linux swap system, which cannot proactively reclaim cold memory, neither saves memory nor degrades performance. The potential performance of a proactive reclamation system using the Linux kernel swap mechanism is evaluated separately in~\cref{sec:eval-kernel-baseline}.

We evaluate the following long-running cloud benchmarks: 

\begin{description}[leftmargin=0em,labelindent=-0.5em,noitemsep]
    \item{\textbf{bert}}~\cite{bert} is a CPU-based ML inference workload using the BERT-Large model and the reference implementation of mlperf~\cite{mlperf_inference}. As performance metrics we use the average of the inference latency when querying one sample per second.
    \item{\textbf{XSBench}}~\cite{xsbench} is a key computational kernel of the Monte Carlo neutron transport algorithm. We use the $\texttt{runtime}^{-1}$ of this benchmark under 450M lookups in event mode as performance metric.
    
    \item{\textbf{elastic}}~\cite{xsbench} Running Elasticsearch's~\cite{elasticsearch_rally} benchmark tool Rally using 27 different tracks. We use the aggregated query throughput as performance metric.

    \item{\textbf{g500}}~\cite{graph500} The graph500 reference implementation. We use scale $27$ (peak memory consumption around 80GB) and a $16$ vCPUs VM. To decrease the execution time, we set the benchmark to do 2 BFS and 2 SSSP phases. We use $\texttt{runtime}^{-1}$ as performance metric.
    
    \item{\textbf{kafka}}~\cite{kafka} using the included \textit{perf-test}  and report throughput as performance metric.
    
    \item{\textbf{matmul}} Matrix multiplication using OpenBLAS~\cite{openblas} dgemm. We use double precision matrices of size $20480x20480$ for two total iterations on a $4$ vCPUs VM. We use $\texttt{runtime}^{-1}$ as performance metric.
    
    \item{\textbf{nginx}}~\cite{nginx} is benchmarked using the included \textit{wrk} benchmarking tool, using a set of static file serving of 30000 4kB and 5000 1MB files. The peak VM memory utilization is around 9GB. The performance metric used is throughput as reported by \textit{wrk}.
    
    \item{\textbf{redis}}~\cite{redis} is benchmarked using \textit{memtier}~\cite{memtier}. The database is initialized with a 12GB dataset using small keys and 1kB data entries. After initializing the database, a mix of access patterns are executed in sequence: Gauss, Random, Sequential.
\end{description}

Workloads that use a client/server architecture (nginx, redis, kafka) use a dedicated load generator VM. We consider only memory utilization of the VM under test and ensure the load generator VM has enough CPU and memory resources.

\begin{figure}
    \centering
    \includegraphics[trim={0em 1.2em 0em 0.1em},clip,width=0.9\columnwidth]{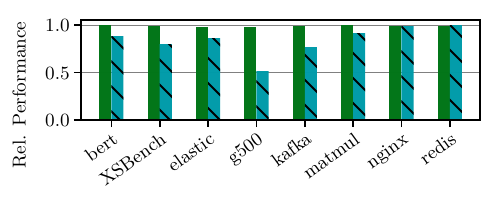}
    \includegraphics[trim={0em 1.5em 0em .8em},clip,width=0.9\columnwidth]{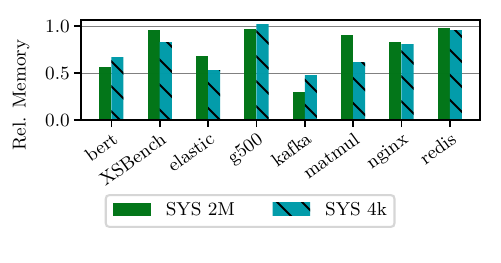}
    \caption{Performance and memory usage}
    \label{fig:eval-softlimit}
\end{figure}

\Cref{fig:eval-softlimit} compares performance and memory saved to a no-swapping system. 2M maintains the application level performance while saving a considerable amount of memory, up to $71\%$ of memory can be reclaimed (Kafka).

Both 2M and 4k save a comparable amount of memory, indicating that most of the 4kB segments of a 2MB page have a similar hotness. Note, that the default reclaimer is a dynamic system with a feedback loop, thus we cannot not expect the exact same amount of memory saved in both 4KB and 2MB case. 

We note that 2M outperforms 4k in all the evaluated workloads. This correlates to our analysis in~\cref{sec:anal-hugepages}, as we corroborate that all of our evaluated workloads in this experiment exhibit a \textit{cold-access percentage} below our cutoff point of $0.0001\%$. It does however not explain the full extent of 4k's slowdown. There are two other factors involved, the \textit{first-touch latency} and \textit{locality}.

The increased \textit{first-touch latency} affects the performance of 4k, in particular on g500:
The initial touch of memory causes a MM invocation, with the latency studied in~\cref{sec:eval-mb-kernel-compare}. The cost is negligible with 2MB pages or small VMs, but for large VMs (such as Graph500, 80GB) the additional round trips to MM almost quadruple the duration of the graph creation phase, which corresponds to $34\%$ of the total slowdown. Excluding the graph construction, the relative performance increases to $85\%$, in line with other slowdowns.
The problem is not inherent to 4k, but rather to strict-4kB systems: we measured similar overheads when forcing Linux to use strict 4kB by disabling THP.

Another factor to consider is the \textit{locality} of the workloads. If a workload accesses many 4kB segments of a 2MB page, swapping in the whole 2MB is beneficial. Measuring this factor is not trivial, but - assuming our 2M and 4k implementations follow a similar reclamation pattern - we can compare the number of page faults of $2MB$ and $4kB$ runs as an approximation. We expect the ratio of $\frac{\texttt{4kB-pfs}}{\texttt{2MB-pfs}}$ to be in the range of 1 (low locality, no 4kB reuse) and 512 (high locality, all 4k segments are being reused).

We found that
most workloads have a page fault ratio of close to 500 (kafka, xsbench, bert, graph500). The latter has enough locality to benefit from 2MB pages. Note that the workloads which have a high degree of random accesses, such as Redis, access the memory frequently enough that virtually no reclamation happens at all. We evaluate the performance with forced reclamation in~\cref{ssec:eval-reuse}.

To conclude, in a best-effort reclamation scenario, 2M saves a similar amount of memory as 4k while maintaining performance close to the non-swapping baseline. Our 4k system incurs overheads which are partially attributed to it being a strict-4k system, partially to the developed system's overheads, and, partially to workloads being sensitive to 4kB paging.

\subsection{Our system vs. enhanced Linux reclaim}
\label{sec:eval-kernel-baseline}
\begin{figure}
    \centering
    \includegraphics{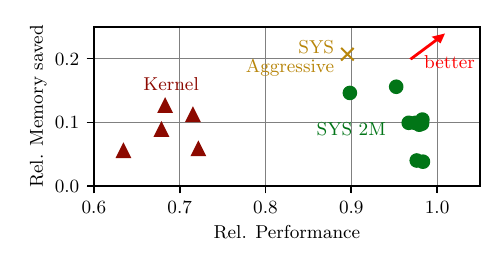}
    \vspace{-10mm}
    \caption{g500 performance and memory usage under different aggressivenes configuration}
    \label{img:g500bl}
\end{figure}

To compare our system's best-effort memory reclamation with a vanilla Linux-based baseline, we port our default reclamation algorithm to control CGroup memory limits and remove our system from the data path.
We modify our EPT scanner (\cref{ssec:impl-ept-scan}) to inform Linux's memory system of pages being young, giving Linux a chance to reclaim unused pages. Linux uses the best-known swapping configuration, THP is enabled (avoiding the first-touch slowdown described in \cref{ssec:eval-softlimit}) and we compensate for QEMU being included in the CGroup (\cref{sec:eval}). 

In \cref{img:g500bl} we report the relative memory savings and performance of 2M and Linux baseline using g500. Runtime and memory consumption are calculated as before (\cref{ssec:eval-softlimit}).
In our initial comparison, we observed that the baseline exhibited poorer performance but achieved greater memory savings. 
We attribute the additional memory savings to the following factors.
First, some access bits from the initial touch are missing in the page table. In the kernel-based implementation, we cannot add these bits because we lack visibility into the page faults. In contrast, our system is able to add the faulting pages to the next access bitmap, making the reclaimer more conservative. Second, unlike the kernel, our system always faults in 2MB pages. If only a subset of the 2MB page is accessed again, the kernel's memory consumption will be lower. Fetching 4kB pages also leads to a degradation of THP coverage. By the end of the benchmark, only 40\% of memory is covered by hugepages, and g500 is sensitive to page size.

To rule out the possibility the performance impact is due to the reclaimer's (\cref{ssec:impl-sys-default-reclaimer}) behavior, we vary the aggressivity of the reclaimer by changing the target swap-in rate and scan interval. We found no configuration where the baseline can match or exceed our system's performance and memory saving.

\subsection{Performance under forced reclamation}
\label{ssec:eval-reuse}

\begin{figure}
    \centering
    \includegraphics[trim={0em 0.7em 0em 0.5em},clip,width=0.9\columnwidth]{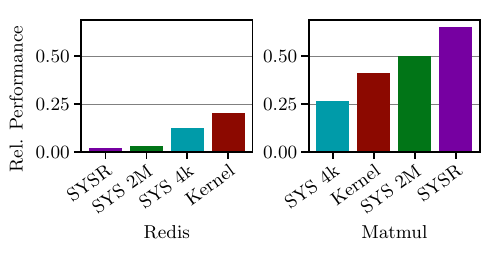}
    \caption{Runtime under $80\%$ memory limit}
    \label{fig:lowmem}
\end{figure}

Setting a memory limit that is significantly below the working set size will heavily impact performance, no matter the system.
We expect a datacenter wide control plane to resolve this situation in a minute timescale.
Nevertheless, such a state can occur transiently and we compare the developed system's performance against Linux swapping.

The most important predictor of workload performance under forced reclamation is determined by its memory access locality. High locality will improve the performance of 2MB swapping, while low locality will improve the performance of 4kB swapping systems. Thus, we compare a workload with a low degree of locality (Redis with random key access) to a workload with a high degree of locality (Matmul).

First, we measure the working set size of each of the benchmarks, and limit the available memory to $80\%$ of the working set size (as before, compensated for QEMU overhead). We then measure the performance under the developed systems in both 4kB and 2MB variants and kernel (THP). We will refer to \sys-R, an improved reclaimer, later in this section.

\cref{fig:lowmem} shows the results.  Redis exhibits better performance in 4kB systems, due to its random access pattern. Matmul, on the other hand, has a high degree of memory access locality, and the developed system has an edge over 4kB systems.

We demonstrate the flexibility of the developed system by implementing a \textit{reuse-distance} based reclaimer \cite{keramidas2007cache, shah2022effective}, \sys-R. This reclaimer approximates Bélády’s optimal algorithm by assigning each page in memory an Estimated Reuse Time (\textit{ERT}), initialized to an IP-sampled reuse-distance and counting down as the page ages. Pages are reclaimed by victimizing the page with the highest absolute ERT.

The -R policy is implemented in under 200 lines of code. The policy is trained on page fault events received from the Policy Engine. For each page fault event, \sys-R employs an IP-based predictor to learn the reuse-distance of the faulting page with respect to historical page faults. The faulting page and its corresponding predicted reuse-distance are then entered in an ERT table. When MM Core requests a page to reclaim, the page with the largest remaining absolute ERT in the ERT table is victimized.

\sys-R reduces the Matmul runtime under thrashing by 30\% over the kernel. Compared to default LRU \sys-R in this case contributes to a $44\%$ reduction in page faults. Due to the random key access in Redis, reuse-distances can not be predicted and \sys-R does not improve upon LRU.

\subsection{Linear prefetcher}
\label{ssec:eval-linearpf}
We demonstrate the utility of \texttt{gva\_to\_hva}~(\cref{des:vm-introspection}) by implementing LinearPF, a simple prefetcher that prefetches the next consecutive page on page fault. We compare two versions, one using GVA and the other using HVA.

The HVA version uses the page fault address directly to find the next consecutive page in memory, while the GVA version maintains a per guest-application context, looks up the corresponding GVA address of the page fault and uses this address to find the next consecutive page.

To show the impact of prefetching in the two different address spaces, we use a synthetic sequential-write workload implemented such that all memory is accessed sequentially from the first to the last page, with sufficient time between each memory access to prefetch the following page. We run this workload for 10 iterations under memory limit corresponding to $75\%$ of WSS on a single vCPU VM. In accordance with the experiment described in \cref{ssec:access-patterns-degrade}, we first warm up the VM by running a random memory access process.

The GVA version improves the runtime by $32\%$ over the baseline, approaching the runtime without a memory limit. In total, the GVA version prefetches over $98\%$ of the page faults in a timely manner. The HVA version does not improve the runtime over the baseline at all, and prefetches under $2\%$ of the page faults.

\subsection{Detecting workload phases}
\label{ssec:eval-aggressive}

We demonstrate flexibility by implementing a policy tailored for workloads operating in phases, where each phase has a substantially different working set.
The \textit{aggressive} policy improves memory reclamation speed for such workloads.
When a workload enters a new phase, it will cause page faults, assuming that some of the new working set is swapped out. The policy detects this increase in page fault rate, and enters the \textit{reclaim mode}. %

\begin{figure}
    \centering
    \includegraphics[trim={0em 0.7em 0.7em 0.5em},clip,width=0.9\columnwidth]{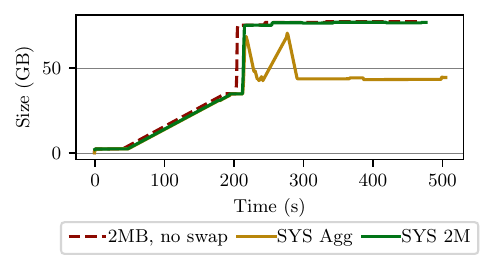}
    \caption{Memory usage of g500}
    \label{fig:graph500-mem}
\end{figure}

Upon entry of the \textit{reclaim mode}, all pages are considered old and are added to an \textit{old page set}. To avoid reclaiming memory used after the uptick, the EPT is scanned every second and removes pages from the old page set that have been accessed. Each second it reclaims and removes up to 2GB from the old page set. When the old page set is empty, the policy leaves the \textit{reclaim mode}.

g500 is an example workload that operates in phases. We compare \sys-Agg with a non-swapping baseline and non optimized system in \cref{img:g500bl}, and we show the memory utilization over time in \cref{fig:graph500-mem}.

From \cref{fig:graph500-mem} we observe that the aggressive policy is able to reclaim memory faster than
the default reclamation policy. Also, increasing the aggressivity of the default reclaimer is not able to save as much memory without drastically sacrificing performance, suggesting that these types of policies are not easily emulated with tuning a default policy.

\subsection{WSR: Lifting memory limit}

\label{ssec:eval-ws-restore}

In this experiment, we consider a transient thrashing state, and investigate the duration of recovering application level performance after.

\begin{figure}
    \includegraphics[trim={1.7em 0.8em 0.0em 0.5em},clip,width=0.9\columnwidth]{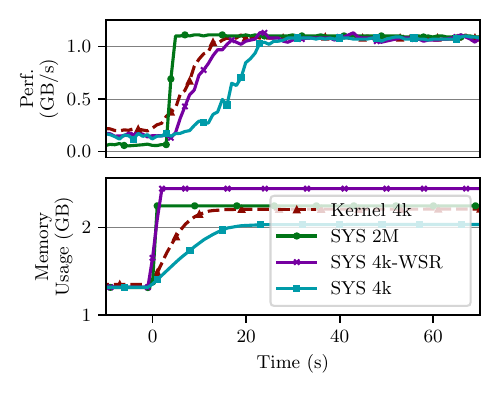}
    \caption{Recovery time after memory limit increase, during memtier benchmark using Redis}
    \label{fig:hardlimit-activewl}
\end{figure}

In \cref{fig:hardlimit-activewl} we compare 2MB, 4k, 4k-WSR (described below) system configurations with Linux swapping. 
We run the memtier benchmark using Redis, and display total memory usage and throughput as performance metric. 2MB and 4k are purely reactive systems, while Linux performs a read-ahead on page faults. The 2MB configuration restores application performance fastest, while 4k takes longest. The kernel, due to its prefetching, is positioned in the middle.

4k-WSR (\textit{working-set-restore}) records the working set, and upon memory limit increase, it prefetches this working set in LRU order. 4k-WSR recovers performance almost as fast as Linux. Prefetching does not map the page, but just removes I/O from the page fault path (it turns major into minor faults). Handling any faults in userspace increases latency, which explains the 4k to kernel gap. 2M, however, is faster than 4kB counterparts because swapping 2MB pages has better I/O throughput.

\section{Related Work}

\mypar{Swapping and virtualization}
vSwapper~\cite{amit_vswapper_2014} introduces optimizations to the swapping mechanism for non-cooperative VM swapping. The optimizations introduced include techniques to reduce unnecessary page-ins and maintain mappings of in-memory pages backed by the VM's virtual disk to reduce disk writes. However, this approach requires significant kernel changes and additional communication channels.

Our system enables user-defined policies and mechanisms for VM swapping, while minimizing kernel modifications. In particular, it allows the implementation of additional optimizations, such as those provided by vSwapper, to be included as part of the userspace mechanism.

\mypar{Cold memory reclamation}
Several works identify cold memory with various techniques and policies, and migrate them to byte-addressable tiered and far memory.~\cite{maruf_tpp_2023,li_pond_2022,kannan_heteroos_2017,park_daos,weiner_tmo_2022,google_far_memory,bergman2022reconsidering,gupta_heterovisor_2015}. However, these approaches necessitate significant modifications to the Linux kernel and resemble a NUMA migration system and not a swapping mechanism.

The developed system also facilitates the detection of cold memory. Additionally, it provides a flexible and user-customizable policy framework, along with easily interchangeable backends, all implemented in userspace.

\mypar{Swapping mechanisms}
Prior works focus on improving the swap performance either by utilizing new NVMe interfaces~\cite{bergman_znswap_2022}, leveraging remote swapping over RDMA~\cite{gu_efficient_2017,amaro_can_2020}, or performing prefetching in the swap subsystem~\cite{maruf_effectively_2020}.

Our system is an extensible userspace swap system and framework, which allows the implemention of the prior approaches in userspace. 

\mypar{Page fault handling in microkernels} Handling page faults in userspace - and optionally by the faulting application itself - has been proposed in various microkernels~\cite{hand1999self, Liedtke95MicrokernelConstruction, bershad1994spin}. We show how these ideas can be implemented on Linux while leveraging a rich ecosystem.

\section{Conclusion}

We present a transparent, flexible, userspace swapping system optimized for virtualization environments. It is designed to better suit cloud environments by leveraging and interacting with commodity software stacks, supporting strict 2MB swapping, and is suitable for opaque VMs.

Our evaluation demonstrate that developed system outperforms existing systems under conservative reclamation, offering significant improvements in performance and memory saving. %

Our system supports custom policies, enabling further performance enhancements tailored to specific workloads. This flexibility ensures that memory management can adapt to diverse cloud environments, providing a robust solution for efficient memory reclamation and utilization.

\onecolumn
\begin{multicols}{2}
\bibliographystyle{ACM-Reference-Format}
\bibliography{main}
\end{multicols}

\end{document}